\documentclass[aps,prd,showpacs,preprint]{revtex4-1}
\usepackage{amsfonts,amsmath,amssymb}
\usepackage{enumerate}
\usepackage{hyperref}
\usepackage{graphicx}
\newcommand{\be}{\begin{equation}}
\newcommand{\ee}{\end{equation}}

\newcommand{\bea}{\begin{eqnarray}}
\newcommand{\eea}{\end{eqnarray}}

\newcommand{\bes}{\begin{subequations}}
\newcommand{\ees}{\end{subequations}}

\newcommand{\nn}{\nonumber}

\newcommand{\ra}{\longrightarrow}
\newcommand{\cN}{{\cal N}}

\newcommand{\Tr}{\mbox{Tr}}

\newcommand{\bgd}{${AdS}_4\times {M}^{1,1,1}$ }
\begin{document}

\title{Rotating Membranes in \bgd}
\date{\today}
\author{Jongwook Kim}
\author{Nakwoo Kim}
\email{nkim@khu.ac.kr}
\author{Jung Hun Lee}
\affiliation{Department of Physics and Research Institute of Basic Science, \\
Kyung Hee University, Seoul 130-701, Korea}

\begin{abstract}
Motivated by the recent progress on gravity duals of supersymmetric
Chern-Simons matter theories,
we consider classical membrane solutions in \bgd. In particular, we present
several types of exact
solutions rotating in the Sasaki-Einstein 7-manifold whose isometry is
$SU(3)\times SU(2)\times U(1)$. We analyze the limiting behavior of macroscopic
membranes and discuss how one can identify the dual operators and the
implications of our result on their conformal dimensions.

\end{abstract}
\maketitle

\pagestyle{plain}
\section{Introduction}
Recently we have seen great progress on our understanding of field theory
duals for various $AdS_4$ backgrounds in string or M-theory.
Most notable is the discovery of ${\cN=6}$ supersymmetric
Chern-Simons-matter theories which provide conformal field theory
duals for $AdS_4\times S^7/\mathbb{Z}_k$ \cite{Aharony:2008ug}.
This three-dimensional field theory has $U(N)\times U(N)$ gauge symmetry
with Chern-Simons (CS) kinetic terms of quantized level $(k,-k)$. The interaction
is also described by a quartic superpotential. In fact, as a quiver guage
theory the data is exactly the same as the well-known conifold theory in four-dimensions
\cite{Klebanov:1998hh},
apart from the extra information on CS levels.

It is an important issue how to generalize this duality to other backgrounds
$AdS_4\times Y_7$. In this paper, we are particularly interested in the examples which
preserve ${\cN=2}$, or eight supercharges. Mathematically $Y_7$ is then
required to be Sasaki-Einstein, and we choose the so-called
$M^{1,1,1}$ space \footnote{Some authors call it $M^{3,2}$.}.
It is constructed as $U(1)$-fibration over
a six-dimensional K\"ahler-Einstein manifold $\mathbb{CP}^2\times\mathbb{CP}^1$.
The dual field theory is thus expected to enjoy $SU(3)\times SU(2)\times U(1)$
global symmetry, where the $U(1)$ part is the usual R-symmetry.

In order to establish the duality relation, we need the Kaluza-Klein
spectrum of 11-dimensional supergravity on $M^{1,1,1}$. It is computed
in \cite{Fabbri:1999mk}, and a three-dimensional quiver gauge theory
was proposed as the dual of $AdS_4\times M^{1,1,1}$ in \cite{Fabbri:1999hw},
but a consistent superpotential could not be written down.
Now with the new insight of Chern-Simons theories without the usual second-order
Maxwell-type kinetic terms, we have a more reliable candidate. The CS duals
have been given for a general class of the so-called $Y^{p,q}(\mathbb{CP}^2)$
metrics \cite{Martelli:2008si,Hanany:2008cd}.
For our interest here the relevant one has gauge symmetry $U(N)^3$ with CS levels $(k,k,-2k)$.
The dual geometry is conjectured to be orbifolds $AdS_4\times M^{1,1,1}/\mathbb{Z}_k$.
A cubic superpotential, with conformal dimension two, is written in terms of the nine bifundamental
chiral multipelts.

The AdS/CFT duality relation implies that the M-theory spectrum in $AdS_4\times M^{1,1,1}/\mathbb{Z}_k$
gives the space of gauge singlet operators on the quiver gauge theory side.
Instead of trying to quantize the supermembrane theory in a curved background, one can study
classical membrane solutions and compare the result to field theory operators with large
conformal dimensions. This is the strategy advocated first in \cite{Gubser:2002tv} for
$AdS_5\times S^5$, and has been extensively used in the study of AdS/CFT relations.
Quantitative results for non-supersymmetric solutions lead to very non-trivial checks of
the duality between the IIB string background and $\cN=4$ super-Yang-Mills theory in
four dimensions. For a review of related works, we refer the readers
to \cite{Beisert:2004ry,Tseytlin:2004xa,Plefka:2005bk}.

The energy and angular momenta of the membrane theory correspond to the conformal
dimension and global charges on the field theory side, respectively. We will consider
various spinning membranes in \bgd background, and when the energy
becomes large, they are expected to give the conformal dimension of very long operators.
Membranes rotating in $AdS_7\times S^4$ are studied in \cite{Alishahiha:2002sy}.
It is also extended to $AdS_4\times S^7$, $AdS_4\times Q^{1,1,1}$ and
various warped backgrounds in \cite{Hartnoll:2002th,Bozhilov:2005xs}.
More papers devoted to exact membrane solutions in $AdS_4\times S^7$ or the IIA background $AdS_4\times \mathbb{CP}^3$ can be
found in \cite{Bozhilov:2005ew,Hoppe:2004iu,
*Bozhilov:2006bi,*Bozhilov:2006gh,*Bozhilov:2007mb,*Wen:2007az,*Bozhilov:2007km,
*Bozhilov:2007wn,*Bozhilov:2007bi,*Ahn:2008gd,*Ahn:2008xm,*Nishioka:2008ib,
*Hamilton:2009iv,*Berenstein:2009sa,*Hamilton:2010sv}.

In order to concentrate on the novelty of the new duality pair,
we will consider membranes moving entirely
in $M^{1,1,1}$. Since the seven-dimensional internal space is described by the scalar fields
via duality, the rotating membranes then correspond to pure-scalar operators. The data of angular
momenta help us identifying the dual operators, and the energy tells us how big is the
anomalous dimension.

This article is organized as follows. In Sec.~\ref{2}, we review the duality proposal for
$AdS_4\times M^{1,1,1}/\mathbb{Z}_k$ and discuss the identification of supersymmetric operators
and Kaluza-Klein modes. In Sec.~\ref{3} we present the gauge fixed membrane action, mainly
to setup the notation. In Sec.~\ref{4.1} we consider particle-like solutions and discuss their
identification as BPS operators. Sec.~\ref{5} is the main part where we present explicit multi-spin
membrane solutions and discuss their field theory duals. We conclude in Sec.~\ref{6}.

\section{M-theory on \bgd and its Chern-Simons dual}
\label{2}
Let us start by presenting the 11 dimensional background \bgd, which preserves
$\tfrac{1}{4}$-supersymmetry.
The metric is given as follows,
\bea
ds^2 &=& ds^2_{AdS_4} + ds^2_{M^{1,1,1}} \, ,
\\
ds^2_{AdS_4}&=& L^2 \left( -\cosh^2 \rho \, dt^2  + d\rho^2 + \sinh^2 \rho
(d\vartheta^2 + \sin^2 \vartheta d\varphi^2 ) \right)
\, ,
\\
ds^2_{M^{1,1,1}}&=&\tfrac{L^2}{64}[d\psi +3\sin^2\mu(d\tilde\psi +\cos\tilde\theta d\tilde\phi)+2\cos\theta d\phi]^2
+\tfrac{L^2}{8}(d\theta^2 +\sin^2\theta d\phi^2)
\nonumber \\
&& +\tfrac{3L^2}{4}[d\mu^2 +\tfrac{1}{4}\sin^2\mu(d\tilde\theta^2 +\sin^2\tilde\theta d\tilde\phi^2 +\cos^2\mu(d\tilde\psi +\cos\tilde\theta d\tilde\phi)^2)] \,.
\label{metric}
\eea
As a configuration of 11 dimensional supergravity, the solution carries 4-form field strength
over $\mbox{Vol}(AdS_4)$. But for the specific type of membrane solutions we will consider,
we can ignore the coupling to gauge field and concentrate on the metric only.

$L$ is the radius of curvature for $AdS_4$, and we have adopted the global coordinate above.
Sasaki-Einstein metric of $M^{1,1,1}$ is here written as a nontrivial
$U(1)$ bundle over $\mathbb{CP}^2
\times \mathbb{CP}^1$. $\theta,\phi$ parametrize $\mathbb{CP}^1$, while
$\mu,\tilde{\psi},\tilde{\theta},\tilde{\phi}$ are for $\mathbb{CP}^2$. Their sizes
are adjusted so that $\mathbb{CP}^2
\times \mathbb{CP}^1$ as a whole becomes K\"ahler-Einstein. The angles range as
$0\leq\theta,\tilde\theta\leq\pi, 0\leq\phi,\tilde\phi\leq 2\pi, 0\leq\psi\leq 4\pi$ and $0\leq\mu\leq \pi/2$.

The above metric for $M^{1,1,1}$ obviously has isometry group $SU(3)\times SU(2)\times U(1)$.
The $U(1)$ part of the Killing vector, $\partial_\psi$, is called the Reeb vector
and accounts for the R-symmetry of the dual field theory. The rest should appear as
additional global symmetry, like flavor symmetry. The ${\cN=2}$ Chern-Simons theory
we will consider in general is dual to orbifold $M^{1,1,1}/\mathbb{Z}_k$, which is
obtained by identifying $\phi \sim \phi + \tfrac{2\pi}{k}$. $SU(2)$ part of the isometry
is then broken to $U(1)$.

\begin{figure}
\includegraphics[scale=0.45,viewport = 200 150 300 350]{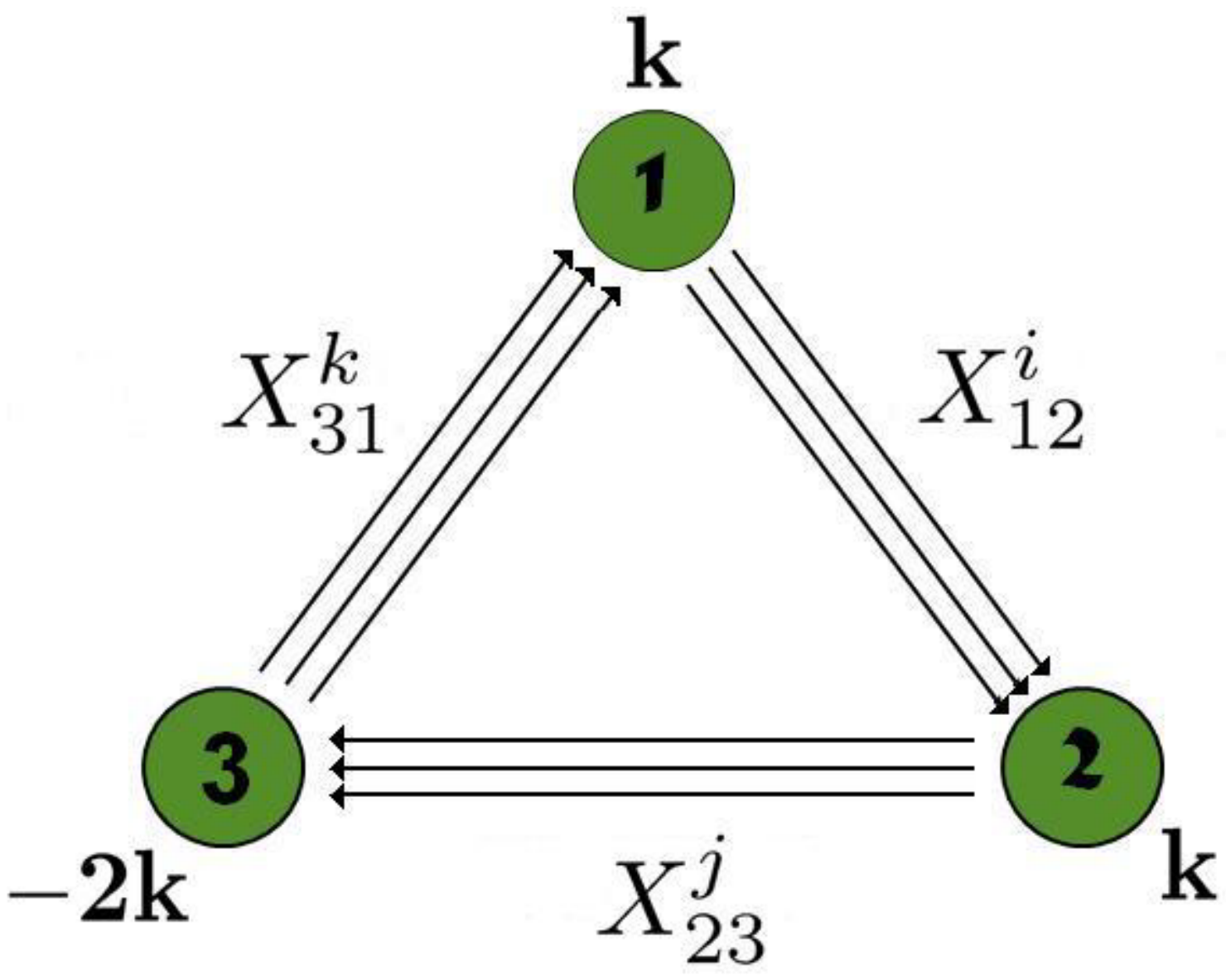}
\caption{\label{fig}Quiver diagram for Chern-Simons dual of \bgd}
\end{figure}

The Chern-Simons dual of \bgd was proposed in
\cite{Martelli:2008si}, and can be summarised by a quiver diagram Fig.~\ref{fig}. We have
$U(N)\times U(N)\times U(N)$ gauge group which is denoted as three nodes in the quiver
diagram. The lines connecting two nodes represent chiral multiplets which are in bifundamental
representations of the two relevant gauge groups. It is essential to note that the CS levels are
not given symmetrically, and take values $(k,k,-2k)$. We also have superpotential,
\be
W = \epsilon_{lmn} \Tr (X^l_{12} X^m_{23} X^n_{31}) .
\ee
The duality can be justified with the help of toric geometry. The mesonic vacuum moduli space,
which is determined by the usual D-term and F-term flatness conditions, is a toric variety and
equivalent to the cone over the orbifold
$M^{1,1,1}/\mathbb{Z}_k$ \cite{Martelli:2008rt,Martelli:2008si,Hanany:2008cd}.
Of course, for $k=1$ the vacuum moduli space has an enhanced isometry group
$SU(3)\times SU(2)\times U(1)$. $X^i_{12},X^i_{23},X^i_{31}$ provide three
fundamental representations of $SU(3)$. The $SU(2)$ broken by the orbifold is
manifest in the quiver diagram.

The spectrum of chiral operators should be consistent with the Kaluza-Klein reduction
of the 11-dimensional supergravity on $M^{1,1,1}/\mathbb{Z}_k$. Chiral operators in
the quiver gauge theory Fig.\ref{fig}, with $k=1$ were studied in Ref.~\cite{Franco:2009sp}.
For pure-scalar
operators, we have a simple task of classifying holomorphic expressions of the nine
complex scalar fields, up to the F-term condition,
\be
\epsilon_{lmn} X^m_{23} X^n_{31} = 0 ,
\quad
\epsilon_{lmn} X^m_{31} X^n_{12} = 0 ,
\quad
\epsilon_{lmn} X^m_{12} X^n_{23} = 0 .
\ee

One can start with the simplest gauge singlet operators,
\be
X^{lmn}_0 = \Tr ( X^l_{12} X^m_{23} X^n_{31} ).
\label{short0}
\ee
There are three $SU(3)$ indices, but the F-term condition forces total symmetrization of
$l,m,n$, so these dimension 2 operators are in ${\bf 10}$ of $SU(3)$.
In order to construct
$SU(2)$ multiplets, we should consider the monopole operators. For simplicity let us
consider the abelian case with CS level $k=1$.
Then there exists a dual photon field $e^{ia}$ which carries charge
$(1,1,-2)$ but does not change the conformal dimension of the entire composite operator. One
can easily verify there are two more neutral operators, and if we continue to
use the matrix notation, they are
\be
X^{lmn}_+ = \Tr (e^{-ia}X^l_{12} X^m_{23} X^n_{23} ), \quad
X^{lmn}_- = \Tr (e^{+ia} X^l_{12} X^m_{31} X^n_{31} ).
\label{short1}
\ee
$X^{lmn}_0,X^{lmn}_\pm$ thus give $({\bf 10}, {\bf 3})$ of $SU(3)\times SU(2)$,
and has
R-charge 2. With generic values of the CS level $k$, $X^{lmn}_\pm$ fields are not
neutral anymore and $SU(2)$ is broken to $U(1)$.

The above spectrum of chiral operators is consistent with the Kaluza-Klein analysis
\cite{Fabbri:1999mk,Fabbri:1999hw}.
As $\cN=2$ supergravity in four dimensions,
there exist hypermultiplets - the ones appearing in Eq.(6.14) of
Ref.~\cite{Fabbri:1999hw} -
which are totally symmetric rank-$3n$ representation of $SU(3)$ and spin-$n$ in $SU(2)$.
Through the AdS/CFT prescription, they have conformal dimension $2n$. In the duality
proposal of Fig.~\ref{fig}, the dual operators are constructed with
$3n$-th order monomials of the scalar fields, and taking the trace.

\section{The membrane action}
\label{3}
For the membrane action, we will use the form developed and utilized by Bozhilov \cite{Bozhilov:2002sj}.
Since we will consider classical solutions which in our ansatz
do not couple to the background 3-form field,
we will just consider the bosonic part and ignore coupling to the gauge field,
for simplicity. For more details see Refs.~\cite{Bozhilov:2005ew,Bozhilov:2002sj}.

We can
start from the
Polyakov-type action
\bea
S^P&=&-\frac{T_2}{2}\int d^{3}\sigma{\sqrt{-\gamma}
(
\gamma^{mn}
G_{mn}
-1)
}\,,
\label{Polyakov}
\eea
where $\gamma_{mn}$ is the auxiliary worldvolume metric and $\gamma=\det \gamma_{mn}$.
$G_{mn}$ is the induced metric,
\be
G_{mn}=\partial_mX^M\partial_nX^Ng_{MN}(X) ,
\ee
where $m,n=0,1,2$ and $M,N=0,1,\cdots,10$. $g_{MN}$ is the background metric, which will
be \bgd for us in this paper.

The variation with respect to $\gamma_{mn}$ leads to constraint
\bea
2G_{mn}-\gamma_{mn}\gamma^{\alpha\beta}G_{\alpha\beta}+\gamma_{mn} =0\,.
\eea
It immediately follows that $\gamma_{mn}=G_{mn}$, and if we plug it back to Eq.~(\ref{Polyakov})
we arrive at the Nambu-Goto action
\bea
S^{NG}=-T_2\int d^{3}\sigma \sqrt{-G}.
\eea
From the Nambu-Goto action,
the generalized momenta are written as
\be
P_M(\sigma)=-T_2 \sqrt{-G}G^{0n}\partial_n X^Ng_{MN} .
\ee
One can then easily check
that $P_M$ satisfy the following constraints ($i=1,2.$)
\be
C_0\equiv g^{MN}P_M P_N+T_2^2GG^{00}=0,
\quad
C_i\equiv P_M\partial_iX^M=0 .
\ee

The canonical Hamiltonian is simply zero, thanks to the constraints.
In order to switch to Hamiltonian description, we thus need to follow Dirac's prescription
and take a linear combination of the first class primary constraints $C_{m}$.
\bea
H=\int d^2\sigma(\lambda^0C_0+\lambda^iC_i)\,. \label{Dirac}
\eea
From the Hamiltonian equation of motion, we have
\bea
\partial_{0}X^M=\frac{\partial H}{\partial P_M}=2\lambda^0 g^{MN}P_N+\lambda^{i}\partial_{i}X^M \nonumber \\
\Rightarrow P_M=\frac{1}{2\lambda^0}g_{MN}(\partial_0-\lambda^i\partial_i)X^N .
\label{momentum}
\eea
We can now obtain an alternative form of Polyakov-type action if we perform the
Legendre transformation once more,
\bea
S^B &=& \int d^{3}\sigma(P_M\partial_0X^M-H) \nonumber \\
    &=& \int \frac{d^{3}\sigma}{4\lambda^0}
\left[G_{00}-2\lambda^jG_{0j}+\lambda^i\lambda^jG_{ij}-(2\lambda^0T_2)^2GG^{00}
\right]
\,.
\eea
Varying this action with respect to $\lambda^k$, we obtain the constraints:
\bea
G_{00}-2\lambda^jG_{oj}+\lambda^i\lambda^j G_{ij}+(2\lambda^0 T_2)^2 \det G_{ij}=0  , \\
G_{0i}-\lambda^iG_{ij}=0 .
\eea
We can work in the gauge, $\lambda^i=0,\lambda^0=\mbox{const}$
\cite{Bozhilov:2005ew}. Then $G_{0i}$=0
and the worldvolume metric takes a block-diagonal form,
and the action is simplified as
\be
S^B =\int \frac{d^{3}\sigma}{4\lambda^0}   [G_{00}-(2\lambda^0T_2)^2 \det G_{ij}]\,.
\label{boz}
\ee
This is the form of the gauge-fixed membrane action we will use in this paper.
The equations of motion derived from Eq.~(\ref{boz}) should be compatible with the
the following constraints,
\bea
G_{00}+(2\lambda^0 T_2)^2 \det G_{ij}=0   \label{constraint2}\,,\\
G_{0i}=0  \label{constraint1}\,.
\eea

\section{Membranes rotating in $M^{1,1,1}$ and their angular momenta}
Let us now present the ansatz we will use for the spinning membrane solutions. In the action
Eq.~(\ref{boz}), the coordinates
of the 11 dimensional spacetime are treated as fields on three dimensional worldvolume coordinates
$\tau,\sigma_1,\sigma_2$.  We will be using the static gauge and identify as $t=\kappa \tau$.
The motion is
in $M^{1,1,1}$ only, so we will set other coordinates of $AdS_4$ to constants,
i.e. $\rho=\vartheta=
\varphi=0$. This setup is consistent with the action and constraints,
and we effectively have membrane motion in $\mathbb{R}\times M^{1,1,1}$, with metric
(apart from the scale factor $L$)
\bea
ds^2&=&
-dt^2
+
\tfrac{1}{64}\left[d\psi +3\sin^2\mu(d\tilde\psi +\cos\tilde\theta d\tilde\phi)+2\cos\theta d\phi
\right]^2
+\tfrac{1}{8}(d\theta^2 +\sin^2\theta d\phi^2)\nonumber \\
&&+\tfrac{3}{4}\left[
d\mu^2 +\tfrac{1}{4}\sin^2\mu\left(d\tilde\theta^2 +\sin^2\tilde\theta d\tilde\phi^2
+\cos^2\mu(d\tilde\psi +\cos\tilde\theta d\tilde\phi)^2\right)
\right]
\,. \label{metric2}
\eea

To describe rotations, we will give linear $\tau$-dependence for {\it azimuthal} angles, and set
\be
\psi=
\zeta\tau, \quad\tilde\psi=\nu_1\tau, \quad\tilde\phi=\nu_2\tau, \quad\phi=\omega\tau .
\label{azi}
\ee
And for {\it polar} angles $\mu,\tilde\theta$ and $\theta$, we assume they do not have
$\tau$-dependence, like $\mu(\sigma_1,\sigma_2)$.

With this choice, we can readily write down several constants of motion.
They are momenta conjugate to those coordinates with linear $\tau$-dependence,
and it is natural to call them $E,J_{\psi},J_{\tilde\psi},J_{\tilde\phi},J_{\phi}$.
It is easy to infer that
$E$ gives conformal dimension, $J_{\psi}$ gives R-charge, and $J_{\phi}$
gives the spin quantum number of $SU(2)$ global symmetry, for the dual operators
in the field theory.

Maybe it is not immediately clear how the $SU(3)$
angular momenta $J_{\tilde\psi}$ and $J_{\tilde\phi}$ give the weight vector of
a given state. Let us follow the the standard convention and
define the Cartan subalgebra with Gell-mann matrices
\be
\lambda_3 = \frac{1}{2} \begin{pmatrix}1 & 0 & 0 \\ 0 & -1 & 0 \\ 0 & 0 & 0  \end{pmatrix}
, \quad
\lambda_8 = \frac{1}{2\sqrt{3}} \begin{pmatrix}1 & 0 & 0 \\ 0 & 1 & 0 \\ 0 & 0 & -2  \end{pmatrix} , 
\label{gel}
\ee
for the fundamental representation. Then
 it turns out that we can relate
$J_{\tilde\phi}$ with $\lambda_3$, and $J_{\tilde\psi}-2J_{\psi}$ with $\lambda_8$.
This adjustment will be justified when we study particle-like solutions.

Below we record the expressions for various conserved quantities of spinning membrane
configurations. 
We have introduced new symbols 
$\Delta\equiv E,
R\equiv 8J_\psi, Q_3\equiv 2J_{\tilde\phi},Q_8\equiv 2\sqrt{3}(J_{\tilde\psi}-2J_{\psi}),
J_3\equiv 2J_\phi$ for later convenience, and $\sqrt{\lambda'}=L^2/2\lambda^0$.
\bea
\Delta&=&\sqrt{\lambda'}\kappa \,, \\
R &=&\frac{\sqrt{\lambda'}}{8}\int \frac{d\sigma^2}{(2\pi)^2}
\left[
\zeta+3\sin^2\mu(\nu_1+\cos\tilde\theta\nu_2)+2\omega\cos\theta
\right] \,,
\label{cd}
\\
Q_3&=&\frac{3\sqrt{\lambda'}}{32}\int \frac{d\sigma^2}{(2\pi)^2}
\left\{
\left[
\zeta+3\sin^2\mu(\nu_1+\cos\tilde\theta\nu_2)+2\omega\cos\theta
\right](\sin^2\mu\cos\tilde\theta) \right.
\nonumber
\\
&&\left.
+4\sin^2\mu
\left[\sin^2\tilde\theta\nu_2+\cos^2\mu(\nu_1+\cos\tilde\theta\nu_2)\cos\tilde\theta
\right]
\right\} \,,
\\
Q_8
&=&\frac{\sqrt{3\lambda'}}{32}\int \frac{d\sigma^2}{(2\pi)^2}
\left\{
\left[\zeta+3\sin^2\mu(\nu_1+\cos\tilde\theta\nu_2)+2\omega\cos\theta
\right](3\sin^2\mu-2)
\right.
\nonumber \\
&&\left.+12\sin^2\mu\cos^2\mu(\nu_1+\cos\tilde\theta\nu_2)\right\} \,,
\\
J_3
&=&\frac{\sqrt{\lambda'}}{16}\int \frac{d\sigma^2}{(2\pi)^2}
\left\{
\left[
\zeta+3\sin^2\mu(\nu_1+\cos\tilde\theta\nu_2)+2\omega\cos\theta
\right]\cos\theta
\right. \nonumber
\\
&& +4\sin^2\theta\omega \Big\}\,.
\label{sp}
\eea

\subsection{Particle-like solutions and their dual operators}
\label{4.1}
We can start by considering rotating particle solutions. When we endow only $\tau$-dependence to the coordinates,
the action and the constraint equations become equivalent to that of geodesic motion. More concretely, Eq.~(\ref{constraint2})
becomes $G_{00}=0$, which is in fact the energy integral of the equations derived from the
action Eq.~(\ref{boz}). Eq.~(\ref{constraint1}) becomes trivial.

One can easily conclude that with uniform rotations as in Eq.(\ref{azi}), the equations for $\mu,\tilde\theta,\theta$
demand that $\mu=0$ or $\tfrac{\pi}{2}$. When $\mu=0$, the value of $\tilde\theta$ is irrelevant,
since the $S^2$ embedded in $\mathbb{CP}^2$ is collapsed. With
$\mu=\tfrac{\pi}{2}$, we can choose $\tilde\theta=0$ or $\pi$. With $\theta$, we always have a choice between
$0$ and $\pi$.

For instance, we can match the different particle-like solutions with vectors in some totally symmetrized tensor representations of
$SU(3)\times SU(2)$ as follows. Let us start with $SU(3)$, and assume the first two entries of the fundamental representations
describe the position in $S^2$ parameterized by $\tilde\theta,\tilde\phi$.
We can then assign
\bea
 \mu=\tfrac{\pi}{2},\,\tilde\theta=0
&\Rightarrow&
e_1
\otimes
e_1
\otimes
\cdots
\otimes
e_1,
\label{111}
\\
 \mu=\tfrac{\pi}{2},\,\tilde\theta=\pi
&\Rightarrow&
e_2
\otimes
e_2
\otimes
\cdots
\otimes
e_2,
\\
 \mu= 0 \quad
&\Rightarrow&
e_3
\otimes
e_3
\otimes
\cdots
\otimes
e_3,
\eea
where $e_1=(1,0,0)^{\rm T},e_2=(0,1,0)^{\rm T},e_3=(0,0,1)^{\rm T}$. The $SU(2)$ part is easy. 
We can interpret orbits at $\theta=\pi \,(0)$
as direct products of spin-up (spin-down) states.

We are now ready to give the dual operators for various particle-like solutions.
For the lowest nontrivial chiral operators, Eqs.~(\ref{short0}) and (\ref{short1})
 explain how they constitute $({\bf 10},{\bf 3})$
representation. It is straightforward to generalize to long operators and match them against
classical point-like solutions. First,
\be
 \mu=\tfrac{\pi}{2},\, \tilde\theta=0, \, \theta=0
 \Rightarrow
 \Tr \left[ ( e^{-ia}X^1_{12} X^1_{23} X^1_{23} )^n \right] .
 \label{111uu}
\ee
This assignment (or the definitions given in Eqs.~(\ref{cd}) to (\ref{sp})) is justified
when we calculate the conserved charges.
As it is usual with supersymmetric
solutions, one can verify that the conserved charges are given by the energy: we find 
$R=\Delta, Q_3=\tfrac{3\Delta}{4},Q_8=\tfrac{\sqrt{3}\Delta}{4},J_3=\tfrac{\Delta}{2}$
from the membrane calculation. And this is precisely what we can just read off from Eq.~(\ref{111uu}), when
we for instance also make use of Eq.~(\ref{gel}) and Eq.~(\ref{111})!

This duality mapping works also with
other solutions: we have
\be
 \mu=\tfrac{\pi}{2},\, \tilde\theta=\pi, \, \theta=0
 \Rightarrow
 \Tr \left[ ( e^{-ia}X^2_{12} X^2_{23} X^2_{23} )^n \right] ,
 \label{222uu}
\ee
with the BPS relation
$R=\Delta, Q_3=-\tfrac{3\Delta}{4},Q_8=\tfrac{\sqrt{3}\Delta}{4},J_3=\tfrac{\Delta}{2}$.
Finally, when we consider membranes located at the south pole of $\mathbb{CP}^2$, we can match
\be
 \mu=0, \, \theta=0
 \Rightarrow
 \Tr \left[ ( e^{ia}X^3_{12} X^3_{23} X^3_{23} )^n \right] ,
 \label{333uu}
\ee
and check $R=\Delta, Q_3=0,Q_8=-\tfrac{\sqrt{3}\Delta}{2},J_3=\tfrac{\Delta}{2}$.
For all the three examples given above, we can switch to lowest weight states of $SU(2)$, by
choosing $\theta=\pi$. Then one needs to change $J_3=-\tfrac{\Delta}{2}$, and
use $e^{ia} X^i_{12} X^i_{31} X^i_{31}, \,\,
i=1,2,3$ for the field theory identification.

\subsection{Membranes extended and rotating in $M^{1,1,1}$}
\label{5}
We now turn to the case where membranes occupy a genuinely three-dimensional worldvolume. 
One first notes that in full generality the membrane action Eq.~(\ref{boz}) is still not easy 
to deal with. It is because of the determinant part $\det G_{ij}$,
which gives rise to terms with four derivatives
in the action. For example  
$\det G_{ij}$ contains terms like
$\tfrac{L^4}{64}
\left[
(\partial_{\sigma_1} \theta)^2 (\partial_{\sigma_2} \theta)^2
-
(\partial_{\sigma_1} \partial_{\sigma_2} \theta)^2
\right]$, and it is obvious that even without the constraints Eq.~(\ref{constraint2}-\ref{constraint1})
the equations of motion are quite nontrivial to handle.

Our strategy to discover nontrivial rotating solutions is
as follows. First, we adopt the temporal gauge and assume 
linear time dependence for $\psi,\tilde\psi,\phi,\tilde\phi$
as in Eq.~(\ref{azi}). Then the constraint
 Eq.~(\ref{constraint1}) is trivially satisfied,
while Eq.~(\ref{constraint2}) is independent of $\tau$. Secondly, 
in order to avoid solving nonlinear partial differential 
equations, we assume each of $\mu,\theta,\tilde\theta$ is
either constant, a linear function of $\sigma_1$, or 
an undetermined function of $\sigma_2$ only. If we also set some of 
the angular velocities $\dot\psi,\dot{\tilde\psi},\dot\phi,
\dot{\tilde\phi}$ to zero, it is possible to make the constraint
Eq.~(\ref{constraint2}) independent of $\sigma_1$, and a function
of $\sigma_2$ only. Then we check if the equations from action
Eq.~(\ref{boz}) is compatible with Eq.~(\ref{constraint2}).
More concretely, in this prescription the equations of motion 
can be treated as classical mechanics system with time variable
$\sigma_2$. Consistency of the ansatz implies Eq.~(\ref{constraint2})
corresponds to the energy integral. 

Our survey has resulted in three physically distinct cases where one 
can consistently reduce the membrane equations of motion into that of an
auxiliary one-dimensional mechanical system. All of them are in general 
extended both in $\mathbb{CP}^2$ and $\mathbb{CP}^1$, and have multiple
non-vanishing angular momenta. We provide the concrete ansatze for 
the angles in the below. We can always give non-vanishing angular velocity for 
$\psi=\zeta \tau$. For other {\it azimuthal} angles like 
$\tilde\psi,\tilde\phi,\phi$, it is needed to set some of their velocities to zero, depending
on the ansatz. We only specified the ones which have to vanish.
\begin{enumerate}[ {Case} 1. ]
\item Rotation in $\mathbb{CP}^2$ and $\mathbb{CP}^1$: $\mu$=$\tfrac{\pi}{2}$, 
$\tilde\theta$=$n\sigma_1$, $\theta(\sigma_2)$; $\nu_2$=0.
\item Rotation in $\mathbb{CP}^2$ only: $\mu$=$\frac{\pi}{2}$, 
$\tilde\theta(\sigma_2)$, $\theta$=$n\sigma_1$; $\omega$=0.
\item Rotation in $\mathbb{CP}^1$ only: $\mu$=$n\sigma_1/2$, $\tilde\theta=0$ or $\pi$,
$\theta(\sigma_2)$;  $\nu_1$=$\nu_2$=0.
\end{enumerate}

For each case, the motion can be described succinctly by the energy integral of an
auxiliary mechanical system. For Case 1, Eq.~(\ref{constraint2}) is reduced to 
\be
\kappa^2 = \frac{3n^2e^2}{128}\theta'^2 + V(\theta), 
\label{en1}
\ee
where $\theta'\equiv\tfrac{d\theta}{d\sigma_2}, e\equiv 2\lambda^0 T_2 L$ and 
the {\it potential} is given as 
\be
V(\theta) = \frac{1}{64}(\zeta+3\nu_1+2\omega\cos\theta)^2+\frac{1}{8}\omega^2\sin^2\theta
\, .
\label{po1}
\ee
We here treat $\sigma_2$ as time, $\kappa^2$ as energy. We find we can set 
$\zeta=0$ or $\nu_1=0$, without losing generality. The potential 
is given in terms of trigonometric functions, so we have a pendulum problem. 
The motion in $\sigma_2$ describes how the membrane is extended along the
{\it longitude} of $\mathbb{CP}^1$. Due to the periodic boundary condition, $n$ is 
integral, and the {\it period} of the motion $\theta(\sigma_2)$ should be $2\pi$.

The reduction to a generalized pendulum motion is very analogous to the study of spinning strings
\cite{Gubser:2002tv,Tseytlin:2004xa}. An oscillatory motion around $\theta=0$ can be 
interpreted as a folded string stretched along the longitude, around the north pole. 
For our membrane, since it is extended
along $\tilde\theta=n\sigma_1$ as well, in total it takes the shape of a {\it cylinder}. 
When $\kappa$ is 
large enough $\theta(\sigma_2)$ enjoys a revolving motion, and the membrane exhibits a 
{\it toroidal} shape. 

For the 3 cases presented above, we have different variables ($\theta$ for Case 1 and 3, $\tilde\theta$ for Case 2) and different masses ($3n^2e^2/64$ for Case 1) for the pendulum. 
If we denote the dynamical variable by $x$, the potential function assumes the following form:
\be
V(x) = (a +b \cos x )^2 + c^2 \sin^2 x \,.
\label{pp}
\ee
For easier reference we provide the essential information for the reduced
membrane equations in Table~\ref{table1}.
\begin{table}[ht]
\caption{Rotating membranes as a pendulum}
\centering
\begin{tabular}{cccccc}
\hline\hline
Type & Variable & Mass & $a$ & $b$ & $c$
\\ [0.5ex]
\hline
 Case 1 & $\theta$ & $\frac{3n^2e^2}{64}$ &$\frac{\zeta+3\nu_1}{8}$ & 
$\frac{\omega}{4}$ & $\frac{\sqrt{2}\omega}{4}$
\\
 Case 2 & $\tilde\theta$ & $\frac{3n^2e^2}{64}$ &$\frac{\zeta+3\nu_1}{8}$ & $\frac{3\nu_2}{8}$ & $\frac{\sqrt{3}\nu_2}{4}$
\\
 Case 3 & $\theta$ & $\frac{3n^2e^2}{256}$ &$\frac{\zeta}{8}$ & $\frac{\omega}{4}$ & 
$\frac{\sqrt{2}\omega}{4}$
\\ [1ex]
\hline
\end{tabular}
\label{table1}
\end{table}

In general the solutions will involve elliptic integrals. In fact the same type of
potentials have appeared in the study of rotating strings in the conifold
\cite{Kim:2003vn,Benvenuti:2008bd}, and one can apply the same mathematical techniques here.
Although the computation is in principle straightforward, 
the   results are  
rather messy for the most general case of Eq.~(\ref{pp}). In this article 
we will only consider two special subclasses: (i) $b=c=0$ and (ii) $a=0$.
Full analysis with Eq.~(\ref{pp}) and extension to more general solutions analogous
to giant magnons \cite{Hofman:2006xt} will be 
reported in a separate publication.

If we set $\omega=0$ in Eq.~(\ref{po1}) the potential becomes a constant and we have
a particularly simple solution, $\theta=m\sigma_2$ with $m\in \mathbb{Z}$. The same 
solutions are obtained if we set $\nu_2=0$ in Case 2.
 Similar but physically distinct
solutions are given from Case 3 if we set $\omega=0$. Both of these solutions uniformly
wrap two {\it polar} angles, and we will call them {\it toroidal rotating membranes}.
Below we will consider such single-spin membranes and provide the dual gauge theory
interpretation. After that, we will consider cases where $a=0$.
\subsubsection{Toroidal rotating membranes}
\begin{itemize}
\item On the equator of $\mathbb{CP}^2$:

For this class of solutions we have $\nu_2=\omega=0$. And without losing generality
we can further set $\nu_1=0$. Among the seven coordinates of $M^{1,1,1}$, we see
that nontrivial ones are $\psi=\zeta\tau,\tilde\theta=n\sigma_1,\theta=m\sigma_2$, 
and $\mu=\tfrac{\pi}{2}$.
One can readily establish the following relationships between conserved charges,
\be
\left(\frac{\Delta}{\sqrt{\lambda}}\right)^2 
= \left(\frac{R}{\sqrt{\lambda}}\right)^2 + \frac{3n^2m^2}{128} ,
\label{di1} 
\ee
where $\lambda=(L^3T_2)^2$, the 't Hooft coupling constant. 
Other angular momenta are given by 
\be
R: Q_3 : Q_8 : J_3 = 1: 0: \frac{\sqrt{3}}{4} R : 0.
\ee
Of course all these quantities
are determined by the angular velocity, for instance $R/\sqrt{\lambda}=\zeta/8$. If we fix 
the values of $n,m$ and increase $\zeta$, the dispersion relation Eq.~(\ref{di1}) saturates
the unitarity bound $\Delta \ge R$. In that sense this type of solution with large $\zeta$ is
{\it near-BPS}. 

The dual operators can be easily inferred using the mapping explained in Sec.~\ref{4.1}.
$R$ basically gives the {\it length} of the operator. $J_3=0$ implies we should put
the same number of $e^{-ia} X^i_{12} X^j_{23} X^k_{23}$ and $e^{ia} X^i_{12} X^j_{31} X^k_{31}$ in the trace. On the other hand, $Q_3=0,\, Q_8=\sqrt{3}R/4$ is realized if
we have half $X^1$ and half $X^2$. For example an operator 
with $R=20$ is written as 
\be
\Tr \left[
(e^{-ia} X^1_{12} X^1_{23} X^1_{23}) (X^2_{12} X^2_{23} X^2_{31})^5
(e^{ia} X^2_{12} X^2_{31} X^2_{31}) (X^1_{12} X^1_{23} X^1_{31})^3 
\right] \,. 
\ee
There are many other gauge singlet operators with the same global charges, and 
the AdS/CFT correspondence predicts there exist eigenstates whose conformal dimension
is given by Eq.~(\ref{di1}). For $SU(3)$, the state is made of $e^1,e^2$ only, so we say
they are {\it along the equator} of $\mathbb{CP}^2$.

\item On the meridian of $\mathbb{CP}^2$:

Here again we set $\nu_1=\nu_2=\omega=0$, and $\mu=n\sigma_1/2, \theta= m\sigma_2$.
We can set either $\tilde\theta=0$ or $\pi$, but for definiteness choose $0$.
Conserved charges are given as follows,
\bea
\left(\frac{\Delta}{\sqrt{\lambda}}\right)^2 
&=& \left(\frac{R}{\sqrt{\lambda}}\right)^2 + \frac{3n^2m^2}{512} \, .
\label{di2} 
\eea
Between the angular momenta, one finds
\be
R: Q_3 :Q_8 : J_3 = 1: \tfrac{3}{8}:  -\tfrac{\sqrt{3}}{8} : 0 .
\label{ra2}
\ee
We can again identify the field theory operators dual to the above classical membrane solution
with large $R$. Eq.~(\ref{ra2}) is realized if
there are half $X^1$ and half $X^3$. If we chose $\tilde\theta=\pi$, 
we would have half $X^2$ and half $X^3$ instead. 
$J_3=0$ again implies we have the same number of
$e^{-ia} X^i_{12} X^j_{23} X^k_{23}$ and $e^{ia} X^i_{12} X^j_{31} X^k_{31}$, in addition to
 an arbitrary fraction of $X^i_{12}X^j_{23}X^k_{31}$. As an illustration we can think of a
dual operator with $R=20$ as follows,
\be
\Tr \left[
(e^{-ia} X^1_{12} X^1_{23} X^1_{23})^2 (X^1_{12} X^1_{23} X^1_{31})^3
(e^{ia} X^3_{12} X^3_{31} X^3_{31})^2 (X^3_{12} X^3_{23} X^3_{31})^3 
\right] \,. 
\ee
The precise identification would again involve diagonalization of the dilatation 
operator $\Delta$ on the field theory side. Since the states are made of $e^1,e^3$ only in 
$SU(3)$, we say they are {\it along the meridian} of $\mathbb{CP}^2$.
\end{itemize}

\subsubsection{Folded/wrapped rotating membranes}
Let us now move to the case of nontrivial potentials. For simplicity, we adjust the 
angular velocities so that the potential takes a relatively simple form,
\be
V(x) = b^2 \cos^2 x + c^2 \sin^2 x . 
\ee
To achieve that 
one sets $\zeta+3\nu_1=0$ for Case 1 and 2, and $\zeta=0$ for Case 3. For the problems at hand,
we always find $b^2<c^2$. 

With energy $\kappa^2$, it is convenient to introduce a parameter $y=\tfrac{\kappa^2-b^2}{c^2-b^2}$
to describe the solutions. If $0<y<1$, the motion is vibrational, while $y\ge 1$ makes it
rotational. As a membrane, a vibrational motion means the worldvolume is folded and only partly 
covers the {\it longitude} along $\theta$ or $\tilde\theta$. In other words, the membrane is 
cylindrical in total. Rotational motion would on the other hand imply a toroidal shape. 

We can express all conserved charges as a function of $y$. The energy $\Delta$ is fixed if
we impose the periodicity condition. For Case 1, the motion is determined by
\be
\kappa^2 = \frac{3n^2e^2}{128} \theta'^2 + \frac{\omega^2}{16} (1+\sin^2 \theta) . 
\label{pp1}
\ee
From $\sigma_2\sim\sigma_2+2\pi$, we have 
\bea
\frac{\pi}{2} &=& \sqrt{\frac{3n^2e^2}{128}}
\int^{\theta_0}_0 \frac{d\theta}{\sqrt{\kappa^2-\frac{\omega^2}{16} (1+\sin^2 \theta)}}
\nn\\
&=& \frac{\sqrt{6}ne}{4\omega} {\bf K}(y),  
\eea
where $y=\tfrac{16\kappa^2}{\omega^2}-1=\sin^2\theta_0$.
${\bf K}$ is the complete elliptic integral of the first kind, and see the appendix
for our convention and some useful formulas.
$\Delta=\sqrt{\lambda'}\kappa$ can be now rewritten as 
\be
\frac{\Delta}{\sqrt{\lambda}}
= \frac{n\sqrt{6(y+1)}}{8\pi} \,{\bf K}(y).
\ee
\begin{table}[t]
\centering
\caption{Conserved charges for cylindrical membranes. $K\equiv{\bf K}(y),
E\equiv{\bf E}(y)$.}
\begin{tabular}{cccc}
\hline\hline
 &  Case 1 & Case 2 & Case 3 \\ 
\hline
$y$ & $\frac{16\kappa^2}{\omega^2}-1$& $\frac{64\kappa^2}{3\nu_2^2}-3$& $\frac{16\kappa^2}{\omega^2}-1$\\
$\Delta$ & $\frac{\sqrt{6(y+1)}}{8\pi}K$ & 
$ \frac{\sqrt{6(y+3)}}{8\pi}K $& 
$\frac{\sqrt{6(y+1)}}{8\pi} K $\\ [0.5ex]
$R$ & $\frac{\sqrt{6}}{16}$ & $\frac{3\sqrt{2}}{16}$ & $\frac{\sqrt{6}}{16}$\\
$Q_3$ & $0$ & 
$\frac{3\sqrt{2}}{32\pi}(4K - E)$ & 
$\frac{3\sqrt{6}}{128}$ \\
$Q_8$ & $\frac{3\sqrt{2}}{64}$ &$\frac{9\sqrt{2}}{64}$& $-\frac{3\sqrt{2}}{128}$\\
$J_3$ & $\frac{\sqrt{6}}{16\pi}(2K- E)$ & $0$ & 
$\frac{\sqrt{6}}{16\pi}(2 K- E)$\\
\hline
\end{tabular}
\label{table2}
\end{table}
One can proceed to calculate all conserved quantities for the three cases. 
We record the results in Table~\ref{table2}. The charges are all normalized by
$\sqrt{\lambda}$, we set $n=1$, and all the elliptic integrals have an argument $y$.

The analysis for toroidal membranes with a nontrivial potential goes in a similar way.
Let us again illustrate it with Case 1. 
From the periodic boundary condition, we have
\bea
\frac{\pi}{2} &=& \sqrt{\frac{3n^2e^2}{128}}
\int^{\pi/2}_0 \frac{d\theta}{\sqrt{\kappa^2-\frac{\omega^2}{16} (1+\sin^2 \theta)}}
\nn\\
&=& \frac{\sqrt{6}ne}{4\omega\sqrt{y}} {\bf K}(1/y) .
\eea
Then the energy is expressed as 
\be
\frac{\Delta}{\sqrt{\lambda}}
= \frac{n\sqrt{6(y+1)/y}}{8\pi} \,{\bf K}(1/y).
\ee
The results for other cases and different charges are reported in Table~\ref{table3}.
One notes that Case 1 and Case 3 give identical results for toroidal membranes.

\begin{table}[t]
\centering
\caption{Conserved charges for toroidal membranes. 
$K'\equiv{\bf K}(1/y), \,E'\equiv{\bf E}(1/y)$.}
\begin{tabular}{ccc}
\hline\hline
 &  Case 1, 3 & Case 2  \\ 
\hline
$y$ & $\frac{16\kappa^2}{\omega^2}-1$& $\frac{64\kappa^2}{3\nu_2^2}-3$
\\
$\Delta$ & 
$\frac{\sqrt{6(y+1)/y}}{8\pi}K'$ & 
$ \frac{\sqrt{6(y+3)/y}}{8\pi} K' $
\\ 
$R$ & $0$ & $0$
\\
$Q_3$ & $0$ & 
$\frac{3\sqrt{2y}}{32\pi}\left(\frac{3+y}{y} K' - E'\right)$ 
\\
$Q_8$ & $0$ & $0$
\\
$J_3$ & $\frac{\sqrt{6y}}{16\pi}\left(\frac{y+1}{y}K'-E'\right)$ & $0$ 
\\
\hline
\end{tabular}
\label{table3}
\end{table}
For both cylindrical and toroidal membranes, 
it is obvious that one can take the large energy limit in $y\rightarrow 1$, where
${\bf K}(y),{\bf K}(1/y)$ develop a logarithmic divergence. For Case 1 and 3, 
the normalized charges scale as 
\be
{\Delta} \approx \sqrt{2} {J_3};
\quad R, Q_3, Q_8 \ra 0 . 
\ee
And for Case 2, 
\be
{\Delta} \approx \frac{2\sqrt{3}}{3} {Q_3};
\quad R, Q_8, J_3 \ra 0 . 
\ee
One can of course do better and expand $\Delta$ in terms of
$J_3$ or $Q_3$, but we relegate the results as well as the
details to the appendix.

Since the unitarity bound $\Delta \ge |R|$ is not saturated
asymptotically, the dual operators 
are not expected to be holomorphic. To help the identification, we quote here
the dictionary of gauge/membrane correspondence. 
\bea
R &=& 2\left[ \# (XXX) - \#(\bar{X}\bar{X}\bar{X}) \right],
\\
Q_3 &=& \tfrac{1}{2} \left[ \# (X^1) +\#(\bar{X}^2) - \# (\bar{X}^1) -\#(X^2) \right] , 
\\
Q_8 &=& \tfrac{1}{2\sqrt{3}} 
\left[ \# (X^1) +\#(X^2) -2 \#(X^3)- \# (\bar{X}^1) -\#(\bar{X}^2)+ 2 \#(\bar{X}^3) \right] , 
\\
J_3 &=& \# (X_{12}X_{23}X_{23}) - \# (X_{12}X_{31}X_{31}) - \# (\bar{X}_{12}\bar{X}_{23}\bar{X}_{23}) 
+ \# (\bar{X}_{12}\bar{X}_{31}\bar{X}_{31}) .
\eea
All we can say about the dual operators is that their composition is such that the vanishing
global charges cancel. For instance, $R\rightarrow 0$ implies we should have the same number of
$X$'s as $\bar{X}$'s for infinitely long operators. 
Thus these states are far from supersymmetric, and it will be highly
nontrivial to verify the correspondence by comparing the spectrum of the quiver Chern-Simons theory.

\section{Discussion}
\label{6}
In this paper we have studied the classical membrane action in a nontrivial 11 dimensional background, \bgd.
The AdS/CFT correspondence implies classical solutions with large energy give an approximate 
description of dual field theory operators which are very long. 
A rigid configuration is required to rotate to satisfy the equation of motion, 
and the angular momenta in the internal
space $M^{1,1,1}$ correspond to some global charges on the field theory side.

We have identified several types of multi-spin membranes, and have shown that the membrane equations 
are conveniently reduced to auxiliary mechanical problems of generalized pendulum. Using the
recently proposed Chern-Simons quiver dual theory, we have provided the candidate dual operators. 
The spectrum of global charges agrees well with the membrane data, and we believe our analysis already
renders strong support on the conjecture proposed in Refs.~\cite{Martelli:2008si,Hanany:2008cd}.

One can think of many avenues to expand from this work. We have identified the pendulum-like
potentials in Table~\ref{table1}, but have not tried the full analysis for the general case of 
nonvanishing $a,b,c$. One can use the techniques employed in Ref.~\cite{Kim:2003vn} and 
obtain implicit relations between the various global charges. 

For the case of $b=c=0$, one obtains a simple dispersion relation like Eq.~(\ref{di1}). 
Relations like $E^2 - J^2 \propto \lambda $ are very well known in the duality between 
$\cN=4$ Yang-Mills and $AdS_5\times S^5$. First obtained from multi-spin strings 
\cite{Frolov:2003qc}, it is well established that the same relation is also derivable 
from integrable spin chains, see Refs.~\cite{Beisert:2004ry,Tseytlin:2004xa,Plefka:2005bk} for 
reviews. It will be very interesting to try to construct the hamiltonian of the dual spin chain model
for \bgd,
which would have $SU(3)\times SU(2)$ symmetry.

One can also look for different types of solutions, or explore different backgrounds. Membrane 
configurations analogous to giant 
magnons \cite{Hofman:2006xt} and spiky strings \cite{Kruczenski:2004wg} 
which might rotate also in $AdS_4$ are interesting subjects. For different backgrounds, 
we have several homogeneous Sasaki-Einstein manifolds whose Chern-Simons dual are
recently proposed. $Q^{1,1,1}$ is a toric Sasaki-Einstein manifold which is $U(1)$ fibration 
over $S^2\times S^2\times S^2$. $V_{5,2}$ is homogeneous but non-toric, and the relevant
M2-brane theory has been investigated in Ref.~\cite{Martelli:2009ga}. We hope to report on 
these problems in the near future.

\acknowledgments
This work was supported by a grant from the Kyung Hee University 
Post-Doctoral fellowship in 2009 (KHU-20090505).

\appendix*
\section{Complete Elliptic integrals and the Nome $q$ expansion}
In this appendix we present the definition and some properties of elliptic integrals
which are needed for the derivation of our result. The complete elliptic integrals of 
the first (${\bf K}$) and the second (${\bf E}$) kind are defined as
\bea
{\bf K} (m) &=& \int^{\pi/2}_0 \frac{d\phi}{\sqrt{1-m \sin^2\phi}} ,
\\
{\bf E} (m) &=& \int^{\pi/2}_0 \sqrt{1-m \sin^2\phi} \, d\phi . 
\eea
For our purpose it is useful to note that 
\\
\bea
\int^{\theta_0}_0 \frac{d\theta}{\sqrt{\sin^2\theta_0-\sin^2\theta}} 
&=& {\bf K} (\sin^2\theta_0) ,
\\
\int^{\theta_0}_0  {\sqrt{\sin^2\theta_0-\sin^2\theta}} \, d\theta
&=& {\bf E}(\sin^2\theta_0) - ( 1 - \sin^2\theta_0) {\bf K} (\sin^2\theta_0) .
\eea
In order to study the elliptic integrals near the logarithmic singularity, it is convenient to
use the q-series, defined as
\bea
q &\equiv & \exp[-\pi{\bf K}(1-m)/{\bf K}(m)]
\\
  &=& \frac{m}{16}+8\left(\frac{m}{16}\right)^2+84\left(\frac{m}{16}\right)^3+\cdots  . 
\eea
Inverting, one obtains
\bea
m=16(q-8q^2+44q^3-192q^4+\cdots).
\eea
Now that we have $m(q)$ and $q(m)$ as given above, we have the following
alternative expansions.
\bea
{\bf K}(m) &=& \frac{\pi}{2}(1+4q+4q^2+4q^4+\cdots) ,
\\
{\bf E}(m) &=& \frac{\pi}{2}(1-4q+20q^2-64q^3+\cdots) .
\eea
And more importantly, 
\bea
{\bf K}(1-m) &=& -\frac{\ln q }{2} (1+4q+4q^2+4q^4+\cdots) ,
\\
{\bf E}(1-m) &=& (1-4q+ 12q^2-32q^3+\cdots) - 4 q \ln q \, ( 1 - 2 q + 8q^2 + \cdots) .
\eea
To get the expansion for ${\bf E}(1-m)$ given above, 
it is convenient to use
the Legendre's relation,
\be
{\bf E}(m){\bf K}(1-m) +
{\bf E}(1-m){\bf K}(m) -
{\bf K}(m){\bf K}(1-m)
 = \tfrac{\pi}{2} . 
\ee

Using the $q$-series expansions we can eliminate the $y$ ($1/y$) 
dependence of conserved charges in Table \ref{table2} (\ref{table3}) and find the functions $\Delta(J_3),\Delta(Q_3)$.
For Case 1 and 3, one can verify
\be
\Delta = \sqrt{2} (J_3 + \tfrac{\sqrt{6}}{16\pi}) 
+ \epsilon \tfrac{\sqrt{3}}{2\pi} \exp \left[
-\tfrac{8\sqrt{6}\pi}{3} (J_3+\tfrac{\sqrt{6}}{16\pi}) 
\right] + \cdots ,
\ee 
and on the other hand for Case 2 
\be
\Delta = \tfrac{2\sqrt{3}}{3} (Q_3 + \tfrac{3\sqrt{2}}{32\pi}) 
+ \epsilon \tfrac{\sqrt{6}}{4\pi} \exp \left[
-\tfrac{8\sqrt{2}\pi}{3} (Q_3+\tfrac{3\sqrt{2}}{32\pi}) 
\right] + \cdots  .
\ee 
In the above, 
$\epsilon$ is $1$($-1$) for toroidal(cylindrical) membranes.

\bibliography{rm}{}

\end{document}